\documentclass[a4paper,11pt]{article}

\usepackage{jheppub} 

\newcommand{\beq}{\begin{equation}}
\newcommand{\eeq}{\end{equation}}
\newcommand{\eq}[1]{(\ref{#1})}
\newcommand{\fig}[1]{Fig.\ref{#1}}

\title{\boldmath Interacting Thermofield Doubles and Critical Behavior in Random Regular  Graphs}

\author[a,b,1]{A. Gorsky,\note{Corresponding author.}}
\author[c]{O.  Valba}

\affiliation[a]{Institute for Information Transmission Problems RAS, 127051 Moscow, Russia }
\affiliation[b]{Moscow Institute for Physics and Technology,Dolgoprudny 141700, Russia}
\affiliation[c]{Department of Applied Mathematics, National Research University Higher School of Economics,\\101000, Moscow, Russia }

\emailAdd{shuragor@mail.ru}
\emailAdd{ovalba@hse.ru}

\abstract{We discuss numerically the non-perturbative effects in  exponential random graphs which are analogue of eigenvalue instantons in  matrix models.
The phase structure of exponential random graphs with chemical potential for 4-cycles $\mu_4$ and degree preserving constraint is clarified. The   first order phase transition at critical value of chemical potential for 4-cycles $\mu_{4}^{RRG}$ 
into bipartite phase with a formation of fixed number of bipartite clusters is found for ensemble of random regular graphs (RRG). We consider the similar  phase transition in combinatorial quantum gravity based of the Ollivier graph curvature for RRG supplemented with hard-core constraint and show that a order of a phase transition at  $\mu_{4}^{cRRG}$  and the structure of emerging phase depend on a vertex degree $d$ in RRG. For $d=3$ the bipartite closed ribbon emerges at $\mu_4>\mu_{4}^{cRRG}$ while for $d>3$ the ensemble of isolated or weakly interacting hypercubes supplemented with the bipartite closed ribbon gets emerged at the first order phase transition with a clear-cut hysteresis. If the additional connectedness condition
is imposed the phase at $\mu_4>\mu_{4}^{cRRG}$ gets identified as the closed chain of weakly coupled hypercubes.
Since the ground state of isolated hypercube is the thermofield double (TFD) we suggest that the dual holographic picture 
involves multiboundary wormholes. Treating RRG  as a model of a Hilbert space for a interacting many-body system we discuss 
the patterns of the Hilbert space fragmentation at the phase transition. We also briefly comment on a possible relation of the found phase transition to the problem of holographic interpretation of a partial deconfinement transition in the gauge theories.}

\begin{document} 
\maketitle
\flushbottom

\section{Introduction}
\label{sec:intro}

Matrix models play a prominent role in a description of  chaotic systems
with large number degrees of freedom. They  govern,for instance, a spectrum of a chaotic
system, the space-time dynamics of some microscopic degrees of freedom, 
discretization of the Riemann surfaces or their moduli spaces \cite{moore}. The
different ensembles of  random matrices correspond to  systems with 
particular symmetry patterns. Spectral density, spectral correlators
and level spacing distribution are among the most popular characteristics
within the matrix model approach.

Large N matrix models enjoy the non-perturbative phenomena - eigenvalue instantons
which are suppressed as $\exp(-N)$ \cite{shenker,david}. The 
partition function for some matrix ensemble reads as
\beq
Z(t_2,t_3 \dots t_n) = \int dM \exp(- \sum_k^n t_k Tr M^k)
\eeq
where the effective  potential for the eigenvalues  in the integration measure 
generically has several extrema. 
In the eigenvalue matrix model the ground state is determined by the
distribution of eigenvalues of $M$ among extrema of effective potential
which depends on the  variables $t_k$. If the term $t_2$
dominates the eigenvalues typically are collected at one extremum and 
the eigenvalue instanton corresponds to the traveling of the single 
eigenvalue from this extremum to another one.
The initial symmetry like $SU(n)$ is broken down
to some subgroup if multiple eigenvalue  instantons are taken into account. Such symmetry
breaking effects are important at the phase transitions at some critical 
values of $t_k$  in the matrix models with  different matrix measures. 
From the physical viewpoint eigenvalue instantons correspond to  baby-universe in 2d gravity, domain walls 
in SYM theory or ZZ branes in Liouville theory.  The review of the 
different manifestations of  eigenvalue instantons can be found in \cite{marino,ambjorn}.

Exponential random graphs are statistical models with the partition function 
\beq
Z(\mu_2,\mu_3 \dots \mu_n) = \sum_{graphs} \exp(- \sum_k^n \mu_k Tr A^k)
\eeq
where $A$ is a graph adjacency matrix and the summation runs over 
some ensemble, say, over  Erd\H{o}s-R\'{e}nyi(ER) ensemble which
involves the potential $TrA^2$ and is 
the network analogue of Gaussian matrix model. 
The sum over  ensembles of graphs substitutes the integration over matrix ensembles in  matrix models.
The terms $\mu_k TrA^k$ in the measure are  similar to $t_k TrM^k$ terms in the measure of matrix models and $\mu_k$ 
provide chemical potentials for number of higher k-cycles in the particular realization.
One can consider the dependence of partition function on the
chemical potentials and look for the phase transitions caused by condensation of
some motifs which are marked by the order parameters $<TrA^k>\neq 0$  \cite{strauss,burda,newman1,newman2,anabele,hovan, 
diaconis,radin1,radin2,metz, coolen1,coolen2}. The condensation of links,
nodes, triangles has been considered and these phase transitions are analogues of the 
familiar criticalities in the conventional matrix models. 
It was shown that the phase transitions are sensitive to the finite-size effects 
\cite{anabele, gorva, coolen2} and the details of the emerging phases  are $N$ dependent 
in several cases.

The additional local constraint imposing strict
degree conservation yields the additional flavor for the condensation phenomena \cite{hovan,coolen1,coolen2}(see
also \cite{foster,tamm} for some earlier observations).
It is in this case the eigenvalue instantons enter the game.
It turns out that degree conservation constraint amounts to the first order   phase transition
with  formation of the multiple weakly interacting clusters which has been 
demonstrated for constrained ER and RRG with  Hamiltonian involving $\mu_3\neq 0$  \cite{hovan}.
The local degree conservation
constraint in the graph ensemble  has a matrix model counterpart as well, it means that
non-singlet sector of a matrix model now matters. 

The phase transition
can be effectively analyzed in terms of network spectrum. In the ER model  with local constraint and RRG
the formation of  network cluster corresponds exactly to the single eigenvalue instanton. Eigenvalue  instantons
upon the averaging over  ensemble form the new non-perturbative soft band in the spectrum of the graph Laplacian. The similar effects 
take place in a spectral density of a
matrix model upon the account of multiple instantons. Soft "`cluster"' band is separated with the gap from the perturbative part of the spectrum. The number of instantons equals to the number of clusters.  Moreover, the spectrum 
of the perturbative band gets deformed at the phase transition and acquires the
triangular shape \cite{hovan} typical for scale-free graphs due to the intercluster interactions.

Following parallels with the matrix models the exponential
random graphs have been also used as the model for the discrete approximation
to quantum gravity in \cite{trug1}. This approach based on the 
Ollivier graph curvature  was nicknamed as combinatorial quantum gravity. 
The graph Ollivier curvature \cite{olivier1,olivier2} gets
reduced to the Ricci curvature in the continuum limit for the large family 
of graph ensembles \cite{trug-kru} including RRG  therefore the Einstein-Hilbert action can be 
recovered. The dimension of emerging Einstein-Hilbert action $D$ is defined by the degree $d$ in RRG  and 
$\mu_4$ upon proper rescaling plays a role of the gravitational coupling. 
It was claimed in \cite{trug1,trug2}
that such model undergoes the second order phase transition into the dense phase at
some $\mu_{4}^c$ without a cluster formation. The $d=3$ example has been 
investigated in \cite{trug3} and the emerging $S^1$ geometry has been identified.

Another popular application of RRG concerns its role 
as a toy model for the Hilbert space  of interacting many-body system.
The RRG mimics the realization of the Hilbert space as the Fock space.
It was  suggested in \cite{kamenev} that  one-particle localization in the Fock
space is related to the localization in the interacting many-body system. It was argued in \cite{basko,mirlin} that indeed   
localization (MBL) in the interacting many-body system  can be mapped to the problem of one-particle localization in the Fock space
with flat diagonal disorder. In \cite{localization} we have treated the one-particle problem in the Fock space 
without the diagonal disorder but with the structural disorder induced by chemical
potential for 3-cycles. It was found that above the phase transition  the states in the non-perturbative
band  get localized while all states in the perturbative band are delocalized.
More recently  RRG with diagonal disorder has been analyzed in \cite{local1,local2,local3,local4,ivan1,local5}, moreover the 
arguments have been presented that RRG plays the role of tricritical point in the space of deformations \cite{local5}.

In this paper we investigate numerically the effects of eigenvalue instantons in the graph
ensembles with degree conservation. Three graph ensembles  are considered: i) RRG ensemble; ii) RRG + hard-core constraint; iii) RRG + hard-core constraint + connectedness constraint.
First we consider numerically RRG perturbed by chemical potential $\mu_4$ for 4-cycles. It is 
found that at $\mu_{4}^{RRG}$ the network gets clusterized however in contrast to the case
of 3-cycles all clusters are bipartite. 
Then we elaborate the model with the additional hard-core constraint for a possible touching of  neighbor 
4-cycles in RRG   suggested in \cite{trug1}  which we denote as cRRG .
The nature of the 
phase transition in cRRG and the  dependence   on  initial configurations
is investigated. We find the first order phase transition 
in cRRG  and  check the
clear-cut hysteresis which confirms the  order of phase transition for $d>3$.
A bit surprisingly it turned out that almost
all emerging bipartite clusters are isolated or weakly interacting hypercubes while the second ingredient of the emerging
clusterized phase for all numerically available degrees  is the composite
bipartite closed ribbon. The number of nodes involved into the closed ribbon 
depends on $\mu_4$. Our finding contradicts the claim
in \cite{trug1,trug2,trug3} about the order of the phase transition for $d>3$
and about the structure of emerging phase.

Finally we introduce the second constraint for cRRG assuming the connectedness of the graph  
and obtain the closed chain of the weakly connected hypercubes at $\mu_4> \mu_{4}^{cRRG}$.
We reproduce the result of \cite{trug3}  for $d=3$ that at $\mu_4 > \mu_{4}^{cRRG}$ the single composite bipartite closed
ribbon emerges. However for $d>3$ we clearly demonstrate that 
the clustering into hypercubes exists  contrary to the claim in \cite{trug1,trug2,trug3}.

The hypercube phase provides the interesting insight on the possible dual holographic 
picture. Indeed it was noted in \cite{ver20} that the hypercube adjacency matrix
exactly coincides with the interaction term in the Maldacena-Qi model of two coupled SYK models \cite{maldacena}.
Hence for a single hypercube we deal with the strong coupling limit of the Maldacena-Qi
model which has thermofield double state (TFD) as the ground state \cite{ver19,ver20}. 
On the other hand the TFD is holographycally 
dual to the eternal black hole \cite{maldacenaold}. The eternal black hole 
was considered as the simple example of the "`geometry from entanglement"'
approach \cite{van09} when the dual classical geometry emerges from 
the entanglement of states of the boundary theory. Here we have the 
highly entangled Majorana dimers which correspond  to the ground state
of individual hypercube.

At  strong coupling phase we obtain the ensemble of hypercubes
which can be completely isolated or weakly interacting. In the holographic picture
it fits with the recent discussion \cite{van20,marolf20} of the holography 
for the ensemble of theories at  multiple bulk boundaries. If 
the components do not interact the corresponding holographic geometry
is assumed to be multiboundary non-traversable wormhole \cite{maoz} which
becomes traversable if the interaction between the components 
is added. The multiple Majorana dimers were also used recently 
for the toy model for holographic quantum error-correcting codes \cite{eisert}.

We shall also make a brief discussion concerning the possible application of our findings
for two more problems.  
In the dimensionally reduced SYM theory down to matrix quantum mechanics the clusterization
of the matrices into the block diagonal form corresponds to the partial breaking of the $SU(N)$ gauge symmetry
down to $SU(M)^k$   symmetry \cite{cotler,hanada1}. We shall compare our findings in RRG with
this interpretation. On the other hand this partial breaking of the gauge symmetry 
presumably corresponds \cite{cotler} to the transforming of a graviton gas  into 
small black holes. We shall also speculate on the relevance of the eigenvalue
instantons for the partial deconfinement phenomena.
Secondly we  make some comments on the possible interpretation of the emerging new objects-
hypercubes and composite bipartite ribbons in the context of Fock space picture
for Hilbert space of interacting many-body system. The phase transition provides 
the different patterns for the fragmentation of the Hilbert space.

The paper is organized as follows. In Section~\ref{sec:crit} we present the results of numerical 
study of RRG perturbed by chemical potential for 4-cycles and supplemented with the different constraints. 
In Section~\ref{sec:holog} the arguments concerning the holographic interpretation of the strong
coupling phase are presented. Section~\ref{sec:fock} is devoted to the aspects  of Fock space
realization of the Hilbert space of many-body system in terms of RRG. The phase transition is considered
as a fragmentation of the Hilbert space.  
The results and the open questions
are collected in Section~\ref{sec:discus}.

\section{Phase transitions in perturbed random regular graphs}
\label{sec:crit}
\subsection{The model description}
Similar to discussion in  \cite{hovan} suppose that we start with some network and rewire links  under the condition that at each step of rewiring we try to maximize the number of 4-cycles $N_{4}$. Which is the equilibrium structure of the entire network? In mathematical terms this question reads as follows. We assign the chemical potential $\mu_4$ to each  4-cycle and  consider the partition function
\beq
Z(\mu_4) = \sum_{\{\rm states\}} \hspace{-0.25cm} {\vphantom{\sum}}' e^{-\mu_4 N_4}
\label{eq:01}
\eeq
where prime in \eq{eq:01} means that the summation runs over all possible configurations of
links ("states"), under the condition of fixed degrees $\{v_1,...,v_N\}$ in all network vertices.

To simulate the rewiring process, one applies the standard Metropolis algorithm with the following rules: i) if under the reconnection the number of 4-cycles is increasing, a move (rewiring)
is accepted, ii) if the number of 4-cycles is decreasing by $\Delta\, N_4$, or remains unchanged, a move is accepted with the probability $e^{-\mu \Delta\, N_4}$. The Metropolis algorithm runs repeatedly for large set of randomly chosen pairs of links, until it converges. In \cite{reconnection} it was proven that such Metropolis algorithm converges to the Gibbs measure $e^{\mu N_4}$ in the equilibrium ensemble of random undirected ER networks with fixed vertex degree or RRG.

In \cite{hovan} it has been shown that given the bond formation probability, $p$, in the initial graph in constrained ER graph or in RRG, the evolving network above the critical value of chemical potential for triangles $\mu_3$  splits into the maximally possible number of clusters, $N_{cl}$:
\beq
N_{cl}=\left.\left [\frac{N}{Np+1} \right]\right|_{N\gg 1}\approx \left [\frac{1}{p}\right],
\label{eq:02}
\eeq
where $[x]$ means the integer part of $x$ and the denominator $(Np+1)$ defines the minimal size of
formed cliques. The asymptotic limit $\sim [p^{-1}]$ at $N\to\infty$ in \eq{eq:02} is independent on the particular set of corresponding vertex degrees, $\{v_1,...,v_N\}$. For RRG  with degree $d$ the number of clusters tends to $[N/(d+1)]$.
It has been shown in \cite{hovan} that clustering of evolving constrained Erd\H{o}s-Renyi network or RRG under condition of 3-cycle maximization, occurs as a first order phase transition where $\mu_3$ is a control parameter. 

\subsection{Clusterization of the perturbed RRG}

\begin{figure}[tbp]
\centerline{\includegraphics[width=12cm]{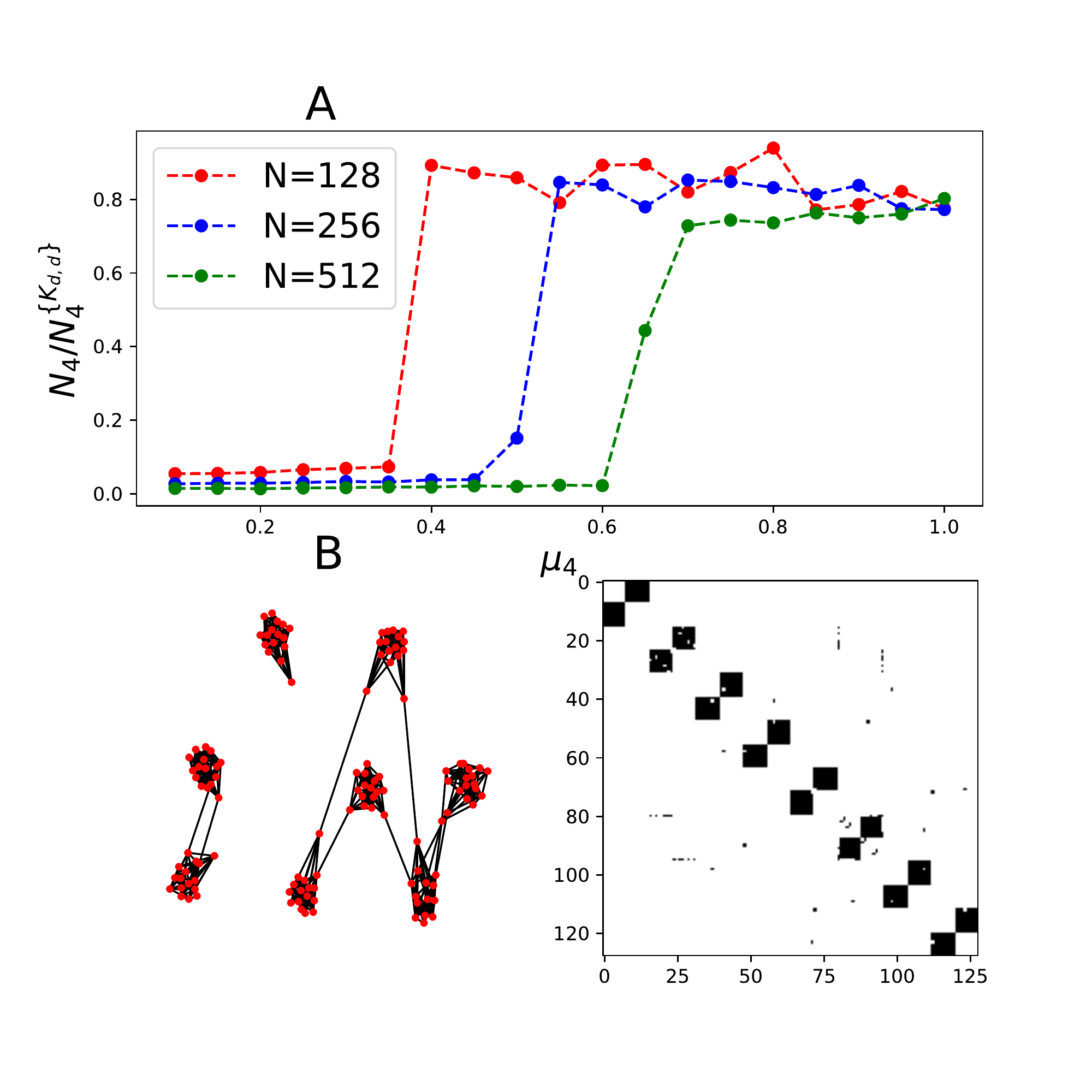}}
\caption{A. The dependence of number of 4-cycles on the chemical potential  $\mu_4$ for RRG of different sizes and $d=8$; B. Ground state and its adjacency matrix at critical $\mu_4$.}
\label{fig01}
\end{figure}

Consider the RRG with degree $d$ perturbed by the $\mu_4$ term
\beq
H=\mu_4 TrA^4
\eeq
and vary the chemical 
potential. At some $\mu_{4,cr}^{RRG}$ the network gets rearranged into bipartite phase with the fixed number
of weakly interacting clusters. The corresponding 
adjacency matrix emerged upon transition is presented at \fig{fig01}. All bipartite 
clusters are almost maximal and have the same size. The similar phase transition takes
place for the constrained ER network however it that case the size of  bipartite clusters
has nontrivial distribution.

\begin{figure}[tbp]
\centerline{\includegraphics[width=16cm]{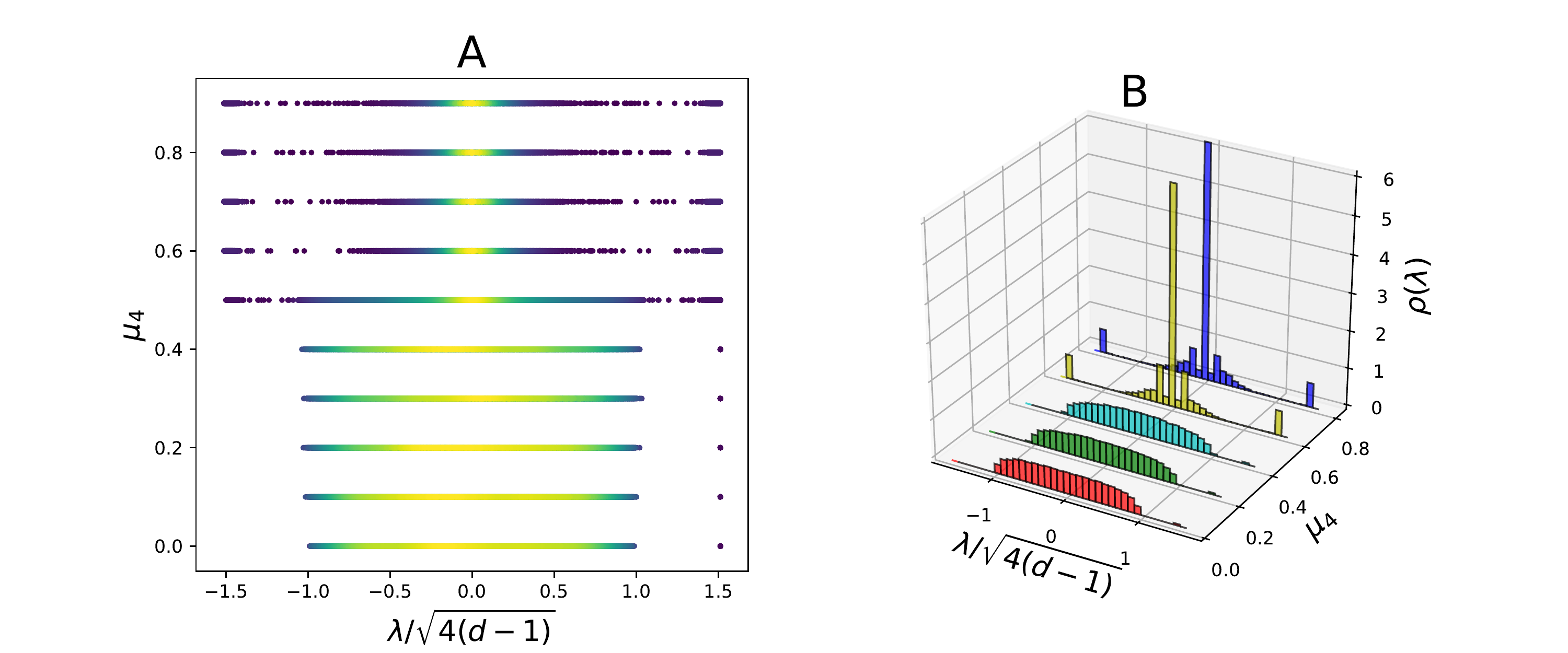}}
\caption{The dependence of the spectral density of adjacency matrix on the chemical potential $\mu_4$ for RRG of $N=256, d=8$.}
\label{fig:02}
\end{figure}

The spectrum of the adjacency matrix $A$ and the Laplacian matrix $L= D-A$ 
where $D=diag(d_1,\dots, d_n)$
is encoded in the spectral density  defined as
\beq
\rho(\lambda)=<\sum_i \delta(\lambda - \lambda_i)>_{RRG}, \qquad Av_i=\lambda_i v_i
\eeq
where the averaging over ensemble is assumed. The unperturbed spectral density
of RRG obeys the Kesten-McKay distribution
\beq
\rho(\lambda)= \frac{\sqrt{4(d-1) -\lambda^2}}{2\pi (d^2 - \lambda^2)}
\eeq

For the RRG a spectral density of the graph Laplacian coincides with
the shifted spectral density of adjacency matrix and we can use both of them
as the Hamiltonian of a particle propagating on the RRG.
The spectral density gets modified upon transition and acquires the following structure.
The central band corresponds to a continuum spectrum. If there would be 
no intercluster interaction there would be 
degenerate states symmetric with respect to the central 
band and degeneracy of eigenvalues would correspond to the number of eigenvalue instantons. We
can interpret these modes as the bound state of the particle localized  at the cluster.
However due to the intercluster interaction we get two non-perturbative
bands instead of the degenerate bound states.
For each realization 
pair of isolated eigenvalues corresponding to the bipartite cluster 
symmetrically tunnel from the perturbative band
in the opposite directions. The evolution of  spectral density of the
adjacency matrix  is presented
at \fig{fig:02}.  

We have also considered the evolution of the spectral density 
for the constrained ER network. The clusterization pattern 
into  bipartite clusters is the same with some peculiarities.
In this case the sizes of the bipartite clusters are different
with some size distribution. The spectral density of graph Laplacian acquires
the three-band structure with more wide non-perturbative bands.

Note that in the spectrum of the graph Laplacian
half of eigenvalues corresponding to  instantons $\lambda_{i,+}$ are 
the soft modes while half of eigenvalues
are the hard modes $\lambda_{i,-}$.  
All  clusters are bipartite hence there is specific interaction 
inside each bipartite cluster which connects different scales corresponding to 
eigenvalues  $\lambda_{i,-},\lambda_{i,+}$. It provides the entanglement
of the bipartite state.

The symmetry breaking pattern at the
phase transition is   $SO(2N)\rightarrow (SO(\frac{N}{d} \times SO(\frac{N}{d})^{d}$.
Let us emphasize that a formation of the bipartite clusters can be observed 
in the real-time during the Metropolis cooling. The formation and evolution of the 
cluster occurs simultaneously with the evolution of pair of isolated eigenvalues. 

\subsection{RRG with hard-core constraint and combinatorial quantum gravity}

Let us consider now cRRG and argue that upon the phase transition
the different ground state emerges.
If only hard-core constraint is imposed the ground state involves the 
ensemble of non-interacting or weakly  interacting hypercubes and a separate bipartite 
closed ribbon. However if additional connectedness condition
is imposed the ground state will be identified as  the closed chain of 
weakly interacting hypercubes.

The hard-core constraint was introduced in the interesting model of combinatorial 
quantum gravity  suggested in \cite{trug1} which is
based on the combinatorial graph curvature invented by Ollivier \cite{olivier1}. The Ollivier curvature
is quite involved for generic graph however it can be simplified
for special types of the graph ensembles. It was suggested in \cite{trug1} to consider
as the simple model for combinatorial quantum gravity the ensemble of RRG with two additional constraints. First, the
bipartiteness condition is imposed and secondly it is assumed that two squares can
have only one common link which to some extent can be considered as the analogue
of the hard-core condition for  extended objects. It was shown recently 
\cite{trug-kru} that for the large class of graphs the Ollivier curvature
gets reduced to the Ricci curvature in the thermodynamic limit.

Under these conditions the density of Ollivier curvature for the cRRG acquires the
simple form
\beq
R_{ij}=-\frac{1}{d}[(2d-2) - N_4(ij)]
\eeq
where $N_4(ij)$ is the number of squares supported at the (ij) edge.
Then one introduces the analogue of the Ricci scalar
\beq
R=\sum_i \sum_j R(ij)
\eeq
which for the cRRG ensemble gets reduced to
\beq
R_{cRRG}= -\frac{8}{d}[\frac{d(d-1)}{2}N - N_4]
\eeq
where $N_4$ is the total number of 4-cycles.

Therefore for cRRG the partition function  involving
Hamiltonian $H=\mu_4 Tr A^4$ acquires the meaning of the combinatorial
version for the action of Euclidean Einstein gravity
\beq
Z_{comb}= \int d g \exp(\alpha \int \sqrt{g} R)
\eeq 
where the integral over the metric comes from the summation 
over the cRRG ensemble.
The chemical potential for 4-cycles defines the gravitational
coupling constant upon some rescaling
\beq
\alpha \propto \mu_4
\eeq

\begin{figure}[tbp]
\centerline{\includegraphics[width=10cm]{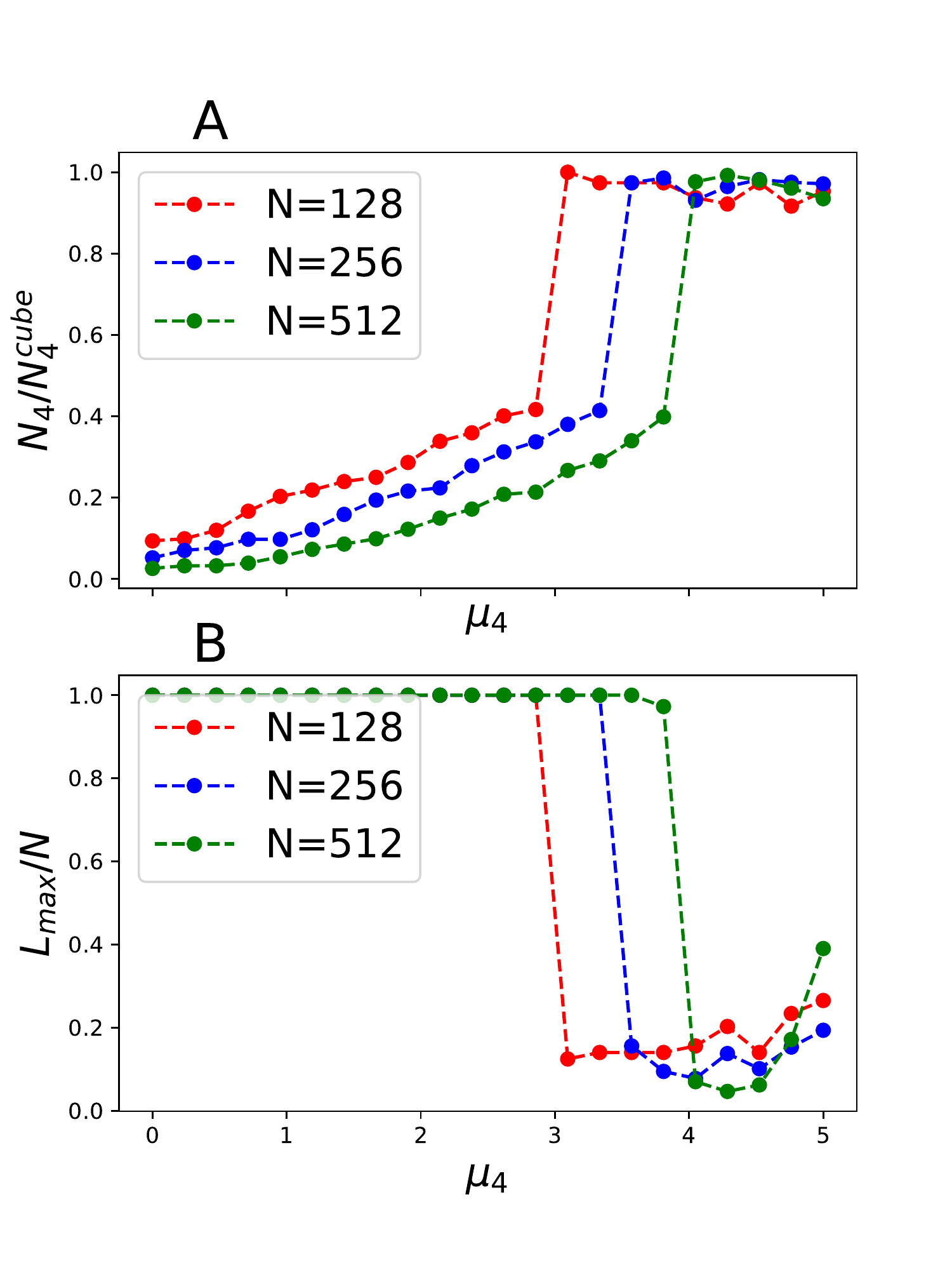}}
\caption{A: The number of 4-cycles in dependence on the parameter $\mu_4$ in cRRG of $d=4$ and different size; B: The portion of the nodes in the bipartite closed ribbon  in dependence on the parameter $\mu_4$.}
\label{fig:03}
\end{figure}

The authors of \cite{trug1,trug2,trug3} claimed that there is the
phase transition at some $\mu_{4}^{cRRG}$ which has the different
nature for RRG and cRRG.  The 
phase transition  in cRRG was assumed to be   the second order
transition without formation of clusters. For the $d=3 $
the emerging phase was found to enjoy $S_1$ topology \cite{trug3}.

\begin{figure}[tbp]
\centerline{\includegraphics[width=16cm]{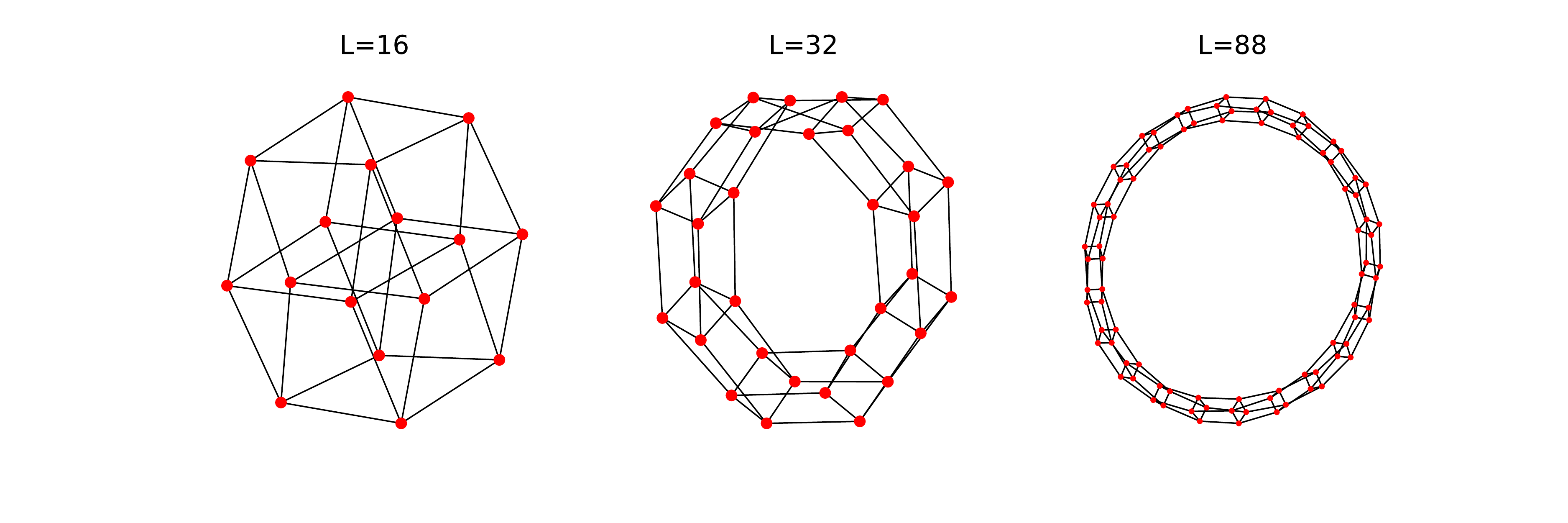}}
\caption{Typical subgraph configurations in the clustered phase involving hypercubes
and composite bipartite strings.}
\label{fig:04}
\end{figure}

We have investigated numerically the phase transition for cRRG for different degrees $d$. First of all, we have found that the first order phase transition takes place at $\mu_{4}^{cRRG}$ for $d>3$. We have observed a set of isolated or weakly connected $d-$ dimensional hypercubes at $\mu=\mu_{4}^{cRRG}$, the number of hypercubes can be estimated as 
\beq
N_{cube}=\left.\left [\frac{N}{2^d} \right]\right|_{N\gg 1}
\label{eq:03}
\eeq
Numerical results for the number of 4-cycles are presented in \fig{fig:03}, there is a explicit hysteresis typical for phase transitions of first order. 
Second, we observe that if there is no enough nodes to form one more hypercube and for $\mu_4>\mu_{4}^{cRRG}$  
part of nodes are organized in the closed bipartite ribbon similar to one discussed in \cite{trug3}. The number 
of the nodes in the closed ribbon depends on $\mu_4$ non-trivially, see \fig{fig:03} B. 
The typical configurations in clusterized cRRG for different values $d$ and $\mu$ are presented at \fig{fig:04}.
Let us emphasize that we do not impose the bipartiteness condition for the initial configurations and 
bipartiteness is emerging symmetry in the clusterized phase. 

\begin{figure}[tbp]
\centerline{\includegraphics[width=16cm]{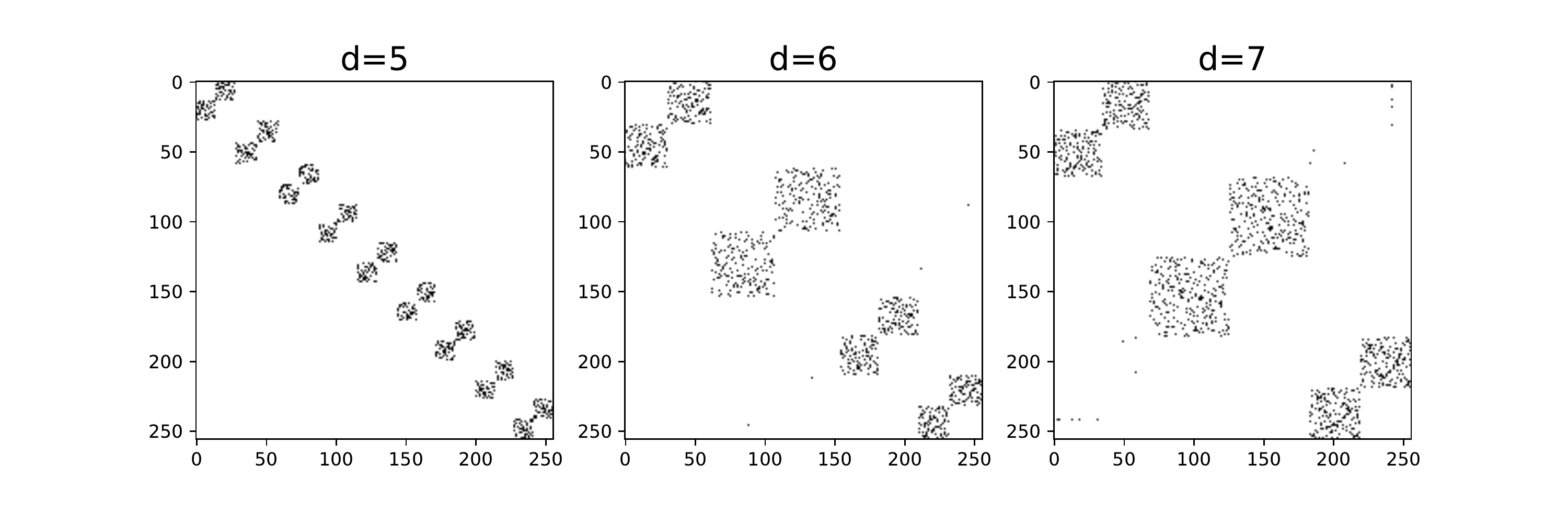}}
\caption{Typical adjacency matrices  in the clustered phase for RRG of $N=256$ and different values of $d$.}
\label{fig:05}
\end{figure}

\begin{figure}[tbp]
\centerline{\includegraphics[width=16cm]{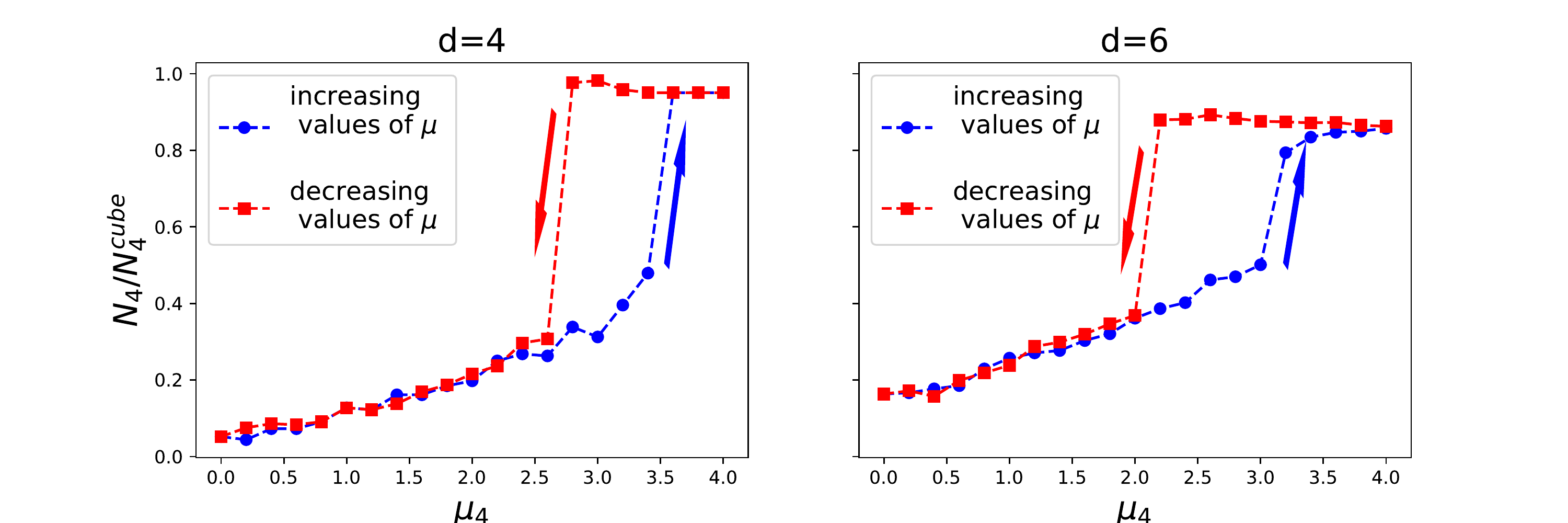}}
\caption{Hysteresis in dependence of the number of 4-cycles for RRG of $N=256$ and different degrees.}
\label{fig:06}
\end{figure}

\begin{figure}[tbp]
\centerline{\includegraphics[width=16cm]{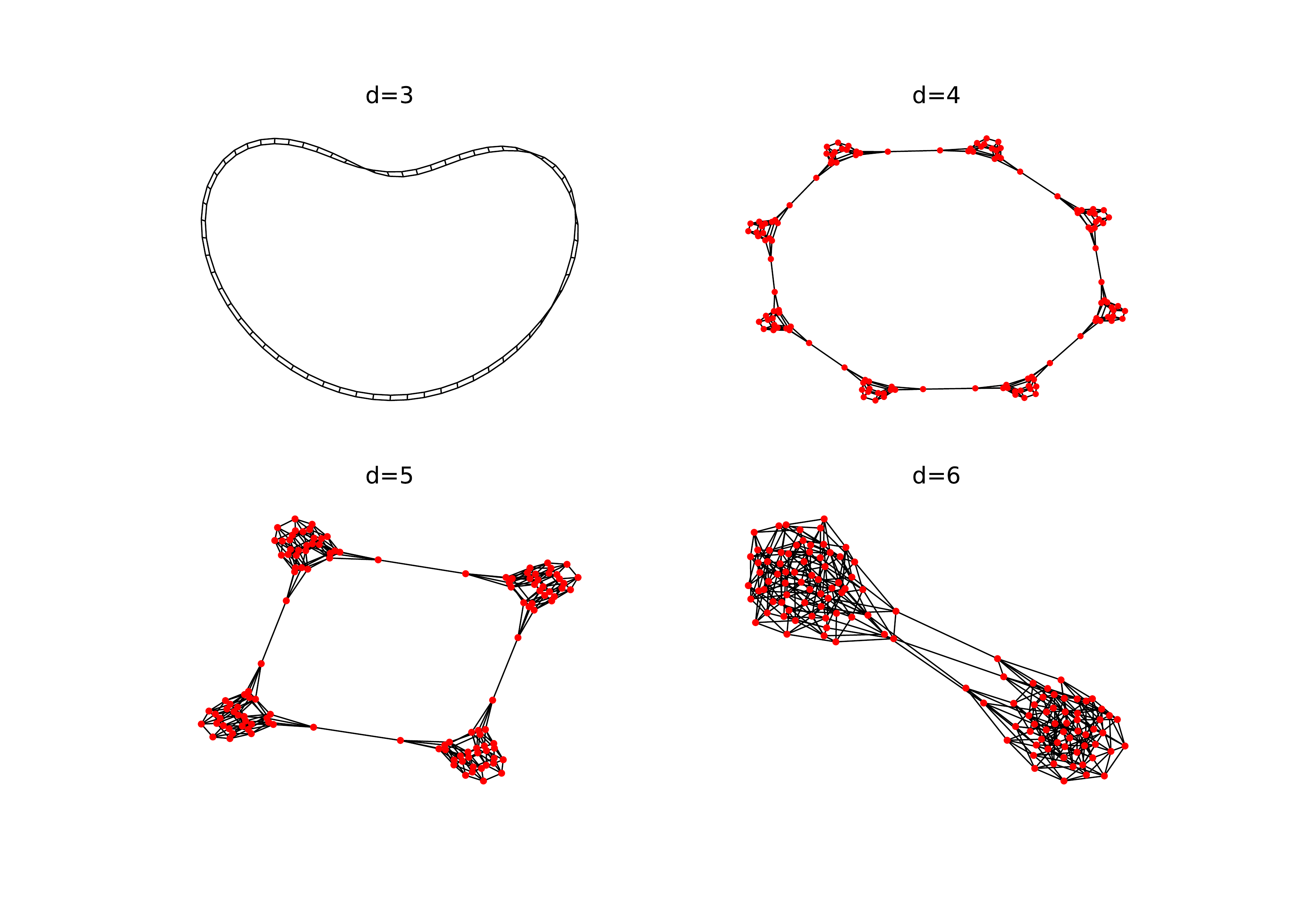}}
\caption{Ground state of cRRG with connectedness constraint and different degree $d$ for RRG of $N=128$.}
\label{fig:07}
\end{figure}

Next, we impose the second connectedness constraint in cRRG and
simulated cRRG with this additional restriction. We have  observed the  phase transition for $d=3$ with the bipartite 
ribbon as the ground state confirming the result of \cite{trug3}. However, for $d>3$ there is a clear-cut  first order phase transition with the formation of the clusterized ground state.
It is identified as closed chain  of weakly connected hypercubes (\fig{fig:07}) and all 4-cycles belong to hypercubes. 
One could wonder why the clusterization in cRRG has not been seen in \cite{trug1,trug2}. It seems, that  the parameters of their numerics did not allow more
than one hypercube hence configuration approximately looks as 
homogeneous.

\section{Interacting Thermofield Doubles and Holographic Dual}
\label{sec:holog}
\subsection{Towards the holographic dual for hypercube phase}

We have argued above that at the transition point the
RRG network with hard-core constraint gets disintegrated into  the ensemble of 
non-interacting or weakly hypercubes with  additional closed bipartite ribbon.
If the connectedness condition is added the emerging configuration is identified as 
the single bipartite ribbon for $d=3$ and the closed chain of weakly connected hypercubes
for $d>3$. In this study we are guided by the  analogy between the
exponential random graphs and matrix models. Hence we could question if a 
kind of holographic representation of our partition function for  the graph ensemble  
can be developed somewhat similar to the matrix model representation of
JT gravity \cite{sss,marolf19}. In that case the $(0+1)$ boundary Euclidean quantum 
mechanics is presented holographically by JT gravity. More precisely
the discussion in \cite{sss} suggest that the bulk gravity is dual 
to the ensemble of boundary theories. 

Before proceeding further two remarks are in order. First, we shall try 
to develop the holographic picture  for the clusterized phase only for 
cRRG and cRRG with connectedness constraint. Secondly, we are guided
by the idea that quantum entanglement of the boundaries can lead to the classical
geometry in the bulk \cite{van09}. The first clear-cut of such situation
has been  developed in \cite{maldacenaold} for the eternal black hole 
with two boundaries. In the Euclidean version the corresponding 
geometry with several boundaries is provided by the multiboundary wormholes \cite{maoz}. The 
boundary components can be entangled via the gravitational bulk. The
recent discussion concerning this scenario can be found in \cite{van20,marolf20}.
We shall have in mind this pattern of holography in what follows.

To develop holographic interpretation we should  clarify what is the meaning of 
hypercubes and bipartite ribbons in the gravity 
framework. First take a look at the hypercubes. It was noted in \cite{ver20}
that the hypercube adjacency matrix $A_{hyper}$ exactly coincides with the interaction
term in the Maldacena-Qi model \cite{maldacena} for the near $AdS_2$ gravity. 
\beq
A_{hyper}= H_{int}=i\sum_k \gamma_k^L \gamma_k^R
\eeq
The mapping is going as follows. Let consider the tensor product representation
for the hypercube Hamiltonian
\beq
H_d =\sigma_0 \otimes H_{d-1} +\sigma_1\otimes\overbrace{\sigma_0\otimes \cdots \otimes
  \sigma_0}^{d-1}, \quad  H_1 =\sigma_1.
\eeq 
and define the $\gamma$ matrices representing Majorana fermions as
\beq
\gamma_k^L = \overbrace{\sigma_1 \otimes\cdots\otimes\sigma_1}^{k-1}\otimes \sigma_3 \otimes \overbrace{\sigma_0 \otimes\cdots\otimes\sigma_0}^{d-k},
\gamma_k^R = \overbrace{\sigma_1 \otimes\cdots\otimes\sigma_1}^{k-1}\otimes \sigma_2\otimes \overbrace{\sigma_0 \otimes\cdots\otimes\sigma_0}^{d-k}
,
\label{gamma}
\eeq
than the Hamiltonian can be written as
\beq
A_{hyper} = i\sum_{k=1}^{d} \gamma_k^L \gamma_k^R.
\eeq

Since we are looking at the RRG ensemble the spectrum of hypercube adjacency matrix which
reads as
\beq
E_n=-d+2n ,n=0\dots d \qquad deg(E_n) = C^d_n
\eeq
coincides with the shifted spectrum of the hypercube graph Laplacian $L=dI-A$ 
where I is identity matrix.
Therefore the one-particle Hamiltonian of a free particle on the hypercube
gets mapped into the interaction term of Maldacena-Qi  Hamiltonian for the Majorana fermions.

On the other hand is was found in \cite{ver19,ver20} that the ground state 
of the Maldacena-Qi model at strong coupling limit  is well approximated by the TFD at $\beta=0$ 
in boundary quantum mechanics of Majorana fermions
\beq
|TFD>= \sum_n e^{-\beta E_n}|n>_{L} |n>_{R}
\eeq
At large N  boundary degrees of freedom in each component of 
TFD are degrees of freedom forming the eternal black 
hole in AdS \cite{maldacenaold}
in Minkowski space or wormhole in the Euclidean space. If there are 
several entangled components the gravity dual is the multiboundary
wormhole.

Now let us turn to our proposal for the gravity dual of the 
clusterized phase. For cRRG we have set of non-interacting 
or weakly interacting hypercubes.
Each hypercube corresponds to the TFD of
entangled  $d$ Majorana fermions. If $d$ is large it is 
natural to assume that each TFD yields the path in the 
dual geometry moreover all TDF's are entangled yielding 
the geometry of non-traversable multiboundary wormhole
in AdS somewhat analogously to the arguments in \cite{van20}
if we have non-interacting but entangled hypercubes. 

If $d$ is small we still have the TFD's of $d$ Majorana fermions
forming the Majorana dimers. It is however impossible to assume
that such TDF corresponds to the fragment of the gravity dual
since it has not enough  degrees of freedom.
On the other hand the total number of TDF,s is large so a kind
of holography can be expected nevertheless. We can use the arguments 
recently suggested in the holographic representation of the 
error-correcting codes \cite{eisert}. It was argued in
\cite{eisert} that boundary Majorana dimers through their
entanglement yield the geodesics in the bulk geometry
allowing its partial restoration. However contrary to 
\cite{eisert} here we have TFD,s for the groups of  
$d$ Majorana dimers.

When the connectedness constraint is added we get
the closed chain of hypercubes and therefore the 
closed chain of TDF,s as ground state. However now
we certainly have the interaction between TDF,s . Following 
arguments from \cite{van20,marolf20} we conjecture that 
the bulk dual is the traversable multiboundary  wormhole 
if $d$ is large enough. At small $d$ we can apply the similar
logic based on the error-correcting codes. Here we have
the interaction between the entangled groups of
Majorana dimers.  

Let us complete this Section with short remark concerning 
the topology of the emerging hypercube phase. The isolated
n-hypercube enjoys the hyperoctahedral symmetry group $C_n$
which is the wreath product of $S_2$ and $S_n$ where 
$S_n$ is the symmetric group of degree $n$. It has nontrivial
first and second homology groups $H_1(C_n,Z) =(Z/2)^2,\quad n\geq 2$,
$H_2(C_n,Z)=(Z/2)^3,\quad n\geq 4$. This allows to look for 
the topologically stable  closed chain of hypercubes.

\subsection{D0-branes quantum mechanics and black hole formation}

Matrix models provide the framework for a formation of extended
objects from ensemble of D0 branes \cite{bfss}. Above we have made the proposal
concerning the holographic interpretation of the
hypercube phase of perturbed cRRG. Let us make a short
comment on the somewhat analogous situation in the 
holographic interpretation of a
matrix model when the typical matrix acquires the block-diagonal form.

According to  holography any transition in the ground
state in the boundary matrix model gets mapped into some 
transition in the dual bulk theory. The transitions can
be identified via the changes of the spectral density and 
the corresponding form of  matrix analogously 
to our discussion of exponential random graphs. It was suggested
in the context of the large N $(1+1)$ SYM theory on the circle that
the distribution of the eigenvalues of Wilson loops plays the key
role. The flat distribution holographically corresponds
to a black string while the gaped distribution corresponds to a 
black hole \cite{aharony}.

Somewhat similar classification has been used in the
context of the deconfinement phase transition \cite{hanada1,berenstein,hanada2,hanada3}.
It was argued that again there are three phases - confined, partially
deconfined and completely deconfined in the boundary theory. The typical matrix 
corresponding to the  partially deconfined phase has deconfined $M\times M$ block
filled densely which presumably corresponds to the small
black hole in the dual bulk while the  rest $SU(N-M)$ sector 
corresponds to the black hole exterior. The entropy in this
phase has intermediate form $S=\epsilon N^2$ where $\epsilon<<1$.
The number of blocks in the typical matrix corresponds to the
number of small black holes.

Having the holographic picture in mind we can speculate on the 
analogy of the phases in the deconfinement transition
with the phases in the exponential random graphs. In the ER network
enriched by 3-cycles  the single dense cluster gets
emerged \cite{strauss} which presumably 
holographically corresponds to the small black hole.
The RRG perturbed by 3-cycles develops several interacting dense
clusters \cite{hovan} which presumably correspond to the   several 
interacting small black holes similarly to \cite{hanada2, hanada3}.

\section{RRG as the model for a Hilbert space and localization}
\label{sec:fock}

The RRG is now the popular model for a investigation of the localization phenomena
in a interacting many-body system. It models the Hilbert space of a model and the MBL in the physical 
space gets mapped into one-body localization at RRG
following the initial arguments suggested in \cite{kamenev}.
The Laplacian of the graph plays the role of the  Hamiltonian
for the propagating degree of freedom.
If the wave function of an effective one-particle system in the Fock space is close to the state
of the initial many-body system, then the localization in the Fock space occurs. Hence, one
can identify the localized state in the Fock space with the particle in the initial many-body
system. Meanwhile, if the wave function of the effective one-particle system in the Fock
space is expanded over a large number of states of the many-body system, this regime is
understood as a delocalized in the Fock space. Similarly to the localization in the real space,
the notion of a mobility edge can be introduced in the Fock space as well.

Technically one can analyze the ergodic properties   through the fractal
dimensions $D_q$, which show  the spread of a state in
the Fock space or via level spacing distribution. Ergodic states at infinite temperatures are 
spread homogeneously over the entire
Fock space hence $D_q = 1$. Instead, non-ergodic states cover
only a vanishing fraction of the Fock space,$0 < D_q < 1$.
while a localization in the Fock space requires $D_q = 0$.
Hence in terms of the Fock space  the
MBL transition  can be identified as transition from $D_q = 1$ in the ergodic space
phase to $D_q < 1$ in the MBL phase.
Different aspects of the one-body localization on RRG with diagonal disorder 
have been discussed in \cite{local1,local2,local3,local4,local5}. Recently
some arguments concerning the possible Kosterlitz-Thouless nature of
this transition in terms of the Fock space have been developed in \cite{new}.

It was argued in \cite{abanin,bernevig}  that 
the so-called scars in the real space could provide the MBL phase. They were
argued to be related to the peculiar fragmentation of the Fock space 
in the many-body system. Generically it is expected that the disorder
is the source of the Hilbert space fragmentation however there are clear examples
\cite{sola,nand} that the local conservation laws also can play such role.
In particular it was shown in \cite{sola,nand} that the combination 
of the charge and dipole number conservation amounts to the 
fragmentation of the Fock space into the exponentially many scars
supporting localized states. In this case there is no need a disorder.
More formal group theory based arguments concerning the fragmentation
of the Hilbert space can be found in \cite{mori,klebanov}. The
use of the Krylov basis for the fragmentation phenomenon has been
discussed in \cite{herviou}.

It was found in \cite{localization} that  the RRG perturbed by the 
chemical potential for 3-cycles $\mu_3$ gets defragmented into the fixed
number of weakly coupled clusters at come critical  $\mu_{3,crit}$. The 
spectrum of graph Laplacian acquires the two-band structure, the number 
of the eigenvalues in the non-perturbative band is equal to the number
of clusters and the eigenvectors in the non-perturbative band are localized.
Hence we have explicit example of the localization in the Fock space via its
fragmentation. One could question about the underlying mechanism 
for the fragmentation in this case similar to the hidden conservation laws in other
examples. The fixed degrees of the nodes in the RRG play the role of
the local conservation laws. To clarify this point we have considered
the constrained ER with the degree conservation constraint. The 
fragmentation of the network similar to the RRG takes place however if
we drop off the degree conservation constraint the pure ER network does not exhibit the
fragmentation phase transition. Note that the degree conservation constraint
in the Fock space
corresponds to the non-local constraint in the real space.

The identification of the graph responsible for the Hilbert space of particular interacting many-body
system is not a simple task. The nodes correspond to the states while the links 
connect the states related by the resonant conditions. The cycles have 
the meaning of the higher resonances \cite{basko2} for instance the 3-cycle
discussed in \cite{localization} corresponds to the three nodes whose energies
are organized in such way that the interaction provides the resonant conditions
for all pairs of states. In this study we have considered the 4-cycles which
correspond to the resonant conditions for four states. We have demonstrated 
that the increasing of the number of four-resonances yields the structural 
disorder in the graph upon the phase transition via the network fragmentation.

We look at the 
localization of the modes in the RRG and cRRG in hypercube phase similarly to \cite{localization}.
Preliminary study shows that similarly to \cite{localization} the modes
corresponding to the bipartite  clusters in RRG are localized in 
agreement with the naive expectations. The pure hypercube corresponds
to the integrable system hence it enjoys the localized modes. However
when the interaction between the hypercubes is switched on it can be
seen that there is localization in the whole spectrum. The question
concerning the existence of mobility edge in this case need for the 
additional study. 

Note that the ER network with chemical potential for 4-cycles
without degree conservation constraint can be considered  as well.
In this case the analytical analysis is available and it perfectly
fits with the numerical simulations \footnote{This result is obtained
in collaboration with A. Vasyliev}. In this case the fragmentation
of the Hilbert space does not take place and all modes are delocalized.
The ER network has the intrinsic disorder in the graph Laplacian
due to the degree distribution however it does not lead
to localization. 

We do not know exactly what physical system has the Hilbert space which we are treating
as the perturbed RRG. However a few remarks are in order. First note that 
the perturbation by 4-cycles amounts to the emerging $Z_2$ symmetry in the
Hilbert space. Secondly one could question if some physical system 
has the hypercube as the Hilbert space. Some indication of the relevance
of the single hypercube to the Hilbert space of  spin  chain has been
noted in \cite{new} but this point certainly needs for further analysis.

Finally note that the Fock space perspective provides the interesting 
twist of the holographic picture. Now we consider the bulk holographic 
dual not for  representation of the boundary partition 
function via the path integral over fields  
but via sum over the  Hilbert space represented by some ensemble of graphs. 
In particular the combinatorial quantum gravity we have discussed
above acquires the meaning of the effective gravity in the Hilbert space
where the chemical potential for the 4-resonances plays the role
of the gravitational coupling. The Ollivier curvature plays the
role of the discrete Ricci curvature in the Hilbert space.
The multiboundary wormhole 
interpretation of the hypercube phase we suggested above 
has the interpretation of the dual bulk geometry for the
boundary Hilbert space ensemble for some theory which we
do not identify.

\section{Discussion}
\label{sec:discus}

In this study we investigated numerically the non-perturbative 
phenomena in  exponential random
graphs with the degree conservation constraints. They are similar to the eigenvalue
instantons in  matrix models however the degree conservation constraint corresponds to the account of the
non-singlet sector in  matrix model context. It was argued that the degree conservation
induces the first order phase transition of a network into bipartite clusters 
at $\mu_4>\mu_{4}^c$ 
both for RRG and constrained ER networks. The induced network bipartiteness at strong coupling 
phase is the dynamical phenomenon.

The RRG with 4-cycle chemical potential has been used as the model  for the combinatorial quantum gravity
if the additional hard-core constraint is imposed. We have performed the detailed numerical  analysis of cRRG  and 
have found some novelties and  disagreements  with results claimed earlier in \cite{trug1,trug2,trug3}.
It turned out that the first order phase transition with a clear-cut  hysteresis  takes place at  $\mu_{4}^{cRRG}$
for $d>3$. 
Nearby the phase transition point the cRRG decays into the 
ensemble of separated non-interacting or weakly interacting  hypercubes supplemented with closed  bipartite ribbon. 
Such constituents of the clustered phase - hypercubes and bipartite ribbon, have appeared
in all numerical experiments at different $d$. 
If we impose the additional graph connectedness condition  for cRRG a structure of the 
clusterized phase gets changed and depends
on value of $d$. At $d=3$ there is a phase transition with the formation of the single 
bipartite closed ribbon in agreement with \cite{trug3}. However for $d>3$ there is  
the first order phase transition with clear-cut
hysteresis and formation of the weakly connected chain of hypercubes contrary to the claims in
the early studies.

Using the relation between the hypercubes and TFD,s we have made the proposal concerning the holographic interpretation
of the bipartite hypercube phase. Namely we have conjectured that the ground state of  cRRG at large $d$ holographically corresponds
to the multiboundary non-traversable wormholes where the number of boundaries coincides with the 
number of isolated hypercubes. At small $d$ when there are no enough degrees of freedom
to form a wormhole we have assumed the relation with a holographic
representation of the error-correcting codes in terms the Majorana dimers \cite{eisert}. The multiboundary
wormhole is supplemented with the bipartite closed ribbon which is related to the 
specific moduli spaces of the mapping of the Riemann surfaces into the sphere with
three branching points \cite{ramgoolam} and presumably deals with the topological membrane. 
If we add the connectedness constraint for cRRG the suggested holographic dual for the
closed chain of weakly interacting hypercubes presumably is
the multiboundary traversable Euclidean wormhole. It would be interesting to relate
this proposal with the tensor network representation of AdS geometry \cite{swingle}
based on the entanglement renormalization. Certainly the holographic dual
of the bipartite hypercube phase deserves a further study.

Our study demonstrates  that non-perturbative phenomena in exponential random graphs 
are more tractable for 
numerical simulations than in the matrix models. This allows to visualize 
the eigenvalue instantons as the process of the creation of the
corresponding  bipartite clusters. In the context of the 
hypercube phase it is also possible numerically to fix $\mu_4$ and increase $N$,
this procedure allows to consider the finite-size effects. Such analysis
in two-star model \cite{gorva} provides the clear picture for the evaporation
of the single cluster. In our case the similar process of a evaporation of the
hypercubes can be expected. It is natural since $N^{-1}$ is the effective coupling. 
Since according to our proposal 
the hypercube at large $d$ being the TFD state is holographically dual to 
the wormhole the evaporation of hypercube
presumably can be related  to a kind of wormhole evaporation. We hope to discuss
this point elsewhere.

Treating RRG as a model for the Hilbert space of many-body system we indicated
a few new patterns of the Hilbert space fragmentation induced by the higher
resonances. The fragmentation of the Hilbert space is related to the
MBL phase in the real space and the preliminary study indicates the one-particle
localization of the cluster modes occurs indeed. We shall present the 
more detailed analysis of the IPR, level spacing distribution and the
entanglement entropy in the hypercube phase elsewhere. It would be also
interesting to consider the RRG supplemented with the fluxes which
would yield the Parisi hypercube model upon fragmentation of the 
Hilbert space. The consideration of the SYK model on the perturbed RRG
along the logic  of \cite{susskind} could be of some importance as well.

One more interesting development concerns the
dual holographic picture for the Hilbert space in the spirit of
the operator-state correspondence. The fragmentation of the
Hilbert space in the hypercube phase provides the firm 
starting point along this line having in mind the wormhole
interpretation of the hypercube ground state.

Our study also provides some suggestions concerning the partial deconfinement scenario. 
In the holographic context the block diagonal form of the matrix presumably corresponds to the 
partial  deconfinement and formation of the small black holes in the  bulk. The 
spectral density of the Wilson or Polyakov loops acquires the gaps. Our study
suggests that the correspondence can be more general and  some version
of wormholes can be relevant. They correspond to the bipartite block diagonal matrices and 
the spectral density with two gaps. It would be interesting to develop
these  arguments further.

Finally remark that the bipartite large $N$ ensemble arises naturally in low energy QCD when the 
integration over  instanton moduli spaces is mimicked by the matrix model
with a simple potential \cite{wettig}. The degree conservation condition 
should be implemented in this model by hand since the 
degree of the node is a topological invariant. 
Our study suggests that when the quartic term in the potential starts to dominate 
in the matrix measure
one could expect that the instanton fluid phase gets transformed 
into the instanton molecular phase when each "` molecule"' corresponds to the
bipartite cluster.


\acknowledgments

We are grateful to I. Khaimovich, V. Kravtsov, A. MIlekhin and S. Nechaev for the useful discussions.
A.G. acknowledge the grant 075-015-2020-801 of Ministry of Science. The work of O.V. is supported within
the framework of the Academic Fund Program at the National Research University Higher School of
Economics (HSE) in 2020–2021 (grant 20-01-041) and within the framework of the Russian Academic
Excellence Project 5-100.
and Higher Education of the Russian Federation.


\end{document}